\documentclass[aps,prl,twocolumn,groupedaddress,showpacs]{revtex4}

\usepackage{dcolumn}
\usepackage{graphicx,amssymb,amsmath}
\usepackage{bm}
\usepackage{natbib}
\usepackage{epstopdf}
\begin{document}

\title{Disorder induced transition into a one-dimensional Wigner glass}

\author{Shimul Akhanjee}
\email[]{shimul@physics.ucla.edu}
\author{Joseph Rudnick}
\email[]{jrudnick@physics.ucla.edu}
\affiliation{Department of Physics, UCLA, Box 951547 Los Angeles, CA 90095-1547}


\date{\today}

\begin{abstract}

The destruction of quasi-long range crystalline order as a consequence of strong disorder effects is shown to accompany the strict localization of all classical plasma modes of one-dimensional Wigner crystals at $T=0$. We construct a phase diagram that relates the structural phase properties of Wigner crystals to a plasmon delocalization transition recently reported. Deep inside the strictly localized phase of the strong disorder regime, we observe ``glass-like'' behavior. However, well into the critical phase with a plasmon mobility edge, the system retains its crystalline composition. We predict that a transition between the two phases occurs at a critical value of the relative disorder strength. This transition has an experimental signature in the AC conductivity as a local maximum of the largest spectral amplitude as function of the relative disorder strength.

\end{abstract}


\maketitle

The behavior of interacting electrons in a random environment has long been a topic of interest in condensed matter physics. When the electrons are confined to a single dimension, many of the complications inherent in the study of such a system are vitiated; a number of studies have focused on the behavior of the collective modes, the various competing manifestations of long-range order and, of course, the equilibrium structural properties of such systems \cite{matveev,schulzwigner,peeters}. More specifically, in the regime in which electronic correlations are expected to be dominant, the Wigner crystal(WC) phase has been proposed as a possible equilibrium configuration in several different 1D and quasi-1D systems\cite{qwire,jap_exp}.

One possible consequence of disorder in any dimensionality is the localization of both electronic wave functions and collective modes \cite{zimandisorder,anderson}. Recently, the plasmons of a disordered 1D WC have been reported to exhibit a delocalization transition \cite{mobility}. Subsequent investigations have clarified the statistical arrangement of the electrons at equilibrium for different types and strengths of disorder \cite{1d_wc}.

\begin{figure}
\centerline{\includegraphics[height=2.0in]{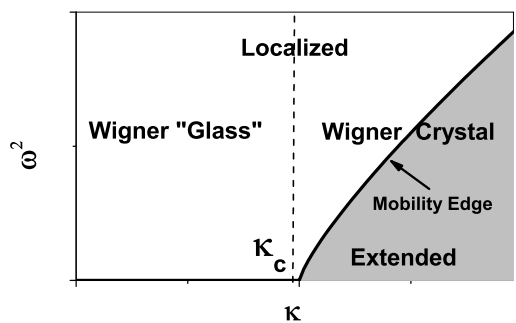}}
\caption{Phase diagram of the disordered 1D Wigner crystal, showing a distinct localized ``glass'' phase for $\kappa < \kappa_c$. The quantity $\kappa$ quantifies the strength of the Coulomb interactions between the localized electrons, in relation to the strength of the disorder (see the discussion below Eq. (\ref{eq:ham})).}
\label{fig:phase_d}
\end{figure}

In many discussions of 1D WC systems at $T=0$ a primary concern is whether the system is truly a crystal in the thermodynamic limit when the effects of quantum-mechanical zero-point motion are included. Previous efforts have shown that---although there may not be true long-range order---the plasmon displacement $u(x)$, correlation functions decay in space much slower than any power law, $\langle {[ {u(x) - u(0)} ]^2 } \rangle  \propto \sqrt {\ln (x)}$. Evidently, this implies that strong Bragg peaks would appear in an experimental probe of the crystal scattering intensities, thus indicating {\em quasi} long-range order\cite{schulzwigner,giampwc}. Furthermore, one is justified in the use of classical methods for studying the plasma modes, due to the lack of particle wave-function overlap at strong unscreened repulsion, existing at lower densities. In this letter we focus on the relationship between the localization of the plasma modes in a one-dimensional array of localized charges (a one-dimensional Wigner Crystal, or 1DWC) and the extent to which quasi-long-range crystalline order prevails in this system in the presence of a ``white noise'' random potential at $T=0$. We first confirm an vital aspect of the plasmon Anderson transition not discussed before, in that at and beyond a critical value of the disorder strength, measured with respect to the interaction magnitude,  all plasma eigenmodes are localized, in analogy to what is seen in the three-dimensional (3D) Anderson transition for non-interacting electrons. Strikingly, we find that all Bragg peaks at $T=0$ also disappear in the immediate neighborhood of this critical value of the disorder. This aspect of the transition can be placed in the context of earlier work on the Anderson transition, in which it is predicted that localization effects tend to stabilize glassy behavior\cite{pastor}. The core of these results are illustrated by the phase diagram shown in Fig.\ref{fig:phase_d}, where the phase boundary separates the frequency $\omega$ regime that the plasma modes of the 1D WC system are localized from the regime in which they are extended.

The physical system is controlled by  the model Hamiltonian of a standard Jellium Wigner Crystal with an added random potential that interacts with each charge \cite{mobility}:
\begin{equation}
H = \sum\limits_{i = 1}^L {\frac{{p_i^2 }}{{2m_e }}}  + \frac{J}{2}\sum\limits_{i \ne j} {\frac{{Q_i Q_j }}{{\left| {x_i  - x_j } \right|}}} + \sum\limits_i^L {Q_i V(x_i)}
\label{eq:ham}
\end{equation}
with
$V(x) = A[\sum\limits_n^N {a_n } \cos (2\pi nx) + b_n \sin (2\pi nx)]$.
The parameters $J$ and $A$ are coupling constants that define the dimensionless interaction strength $\kappa \equiv {J/A} $.  The random variables $a_n$ and $b_n$ are chosen from different Gaussian distributions, yielding a white noise power spectrum with a mean of $\mu=0$ and a variance of $\sigma=1$. The number of Fourier amplitudes, $N$, was set to $N=L/4$ for proper scale invariance. We have observed that the critical mobility edge frequency $\omega^{2}_c$ is generally dependent on the ratio $N/L$ of $V(x)$, which represents the average number of charges available per potential well. The scaling significance of $N$ in finite sized systems is not immediately obvious and must be studied more carefully.

Since we are interested in {\em quasi} long-range crystalline order(due to zero-point motion), it is sufficient to study large finite systems with periodic boundary conditions. The charges were numerically relaxed to their equilibrium configuration with the use of methods are outlined in \cite{1d_wc} and briefly described below. Let us introduce the various physical quantities of interest that are needed to interpret the results of our investigations.

First, the central quantity used for studying the plasma modes is the dynamical matrix $\mathbf{D(R)}$. For a finite sized chain of length $L$, $\mathbf{D(R)}$ is an $L \times L$ symmetric matrix with the structure, $
\mathbf{D(R - R')} = \delta _{\mathbf{R,R'}} \sum_{\mathbf{R''}} {\left. {{\partial ^2 \phi (x)}/{\partial x^2 }} \right|} _{x = \mathbf{R - R''}}
  - \left. {{\partial ^2 \phi (x)}/{\partial x^2 }} \right|_{x = \mathbf{R - R'}} $, where $\phi (x)$ is defined as the electrostatic potential between two charges in a periodic image and the ${\mathbf{R}}$'s are the equilibrium positions of the charges, of which are {\em not} necessarily periodically ordered and must be determined from a numerical equilibration. The eigenvalue equation for plasma eigenmodes follows as \cite{AM}:
$m_e \omega ^2 u(\mathbf{R}) + \sum_{\mathbf{R'}} {\mathbf{D(\mathbf{R - R'})}} u(\mathbf{R'}) = 0$,
with $u(\mathbf{R})$ as the lattice displacements from equilibrium. Apparently, all physical quantities calculated in this article require the determination of the equilibrium ${\mathbf{R}}$'s, therefore an effective numerical relaxation method is absolutely indispensable. We have employed a Newton-Raphson procedure by which one recursively applies the inverse of ${\mathbf{D(\mathbf{R})}}$ to the out-of equilibrium particle coordinates $\mathbf{R}$ until the simultaneous total forces on the charges are sufficiently close to zero\cite{1d_wc} and consequently the system approaches equilibrium.

 An important experimental quantity that follows from the plasma oscillations is the AC conductivity. The plasmon propagator $G(\omega)$, is a measure of the response
of the particle array to an external electric field. $G(\omega)$ can be determined from $\mathbf{D(R)}$, by constructing the resolvent:
$G(\omega ) = \sum_R^L {\sum_{R'}^L {1/{(\omega ^2 \mathbb{I} - D(R,R'))}}}
\label{eq:gfun}$.
Applying the Kubo formula, the AC conductivity is given by \cite{AM}, $\sigma (\omega ) = iA\omega G(\omega )$.

 As in \cite{mobility}, the lengths are scaled so that the size of the region occupied by the charges is unity, $\phi (x) = \pi |\csc (\pi x)|$, and the eigenfunction width is given by $\xi _i  = L^2/{(2\pi ^2 )}\sum_{m,n}^L {u_i (m)^2 u_i (n)^2 \sin ^2 (R_m  - R_n )}$, where $u_i (m)$ and $R_m$ are the respective $ith$ eigenfunction amplitude and particle position at the site $m$.

In terms of the structural properties, we have investigated the integrity of the Fourier space of particle coordinates, given by:
$S_k  = \langle {\sum_{m = 1}^L {e^{ik R_m } } } \rangle /L , \ k = {2\pi m}/L, \ m = 1,2,3 \ldots $
For arbitrarily small disorder strength, the eigenfunctions are no longer pure plane waves. However, some crystalline aspect to the system is conserved, expressed by the standard criterion that there exists at least one Bragg peak in the spectrum of $S_k$. It is, in fact, sufficient to track the behavior of the height of the first Bragg peak, $g(\kappa)$, as the interaction strength $\kappa$ is changed. In the neighborhood of the structural phase transition $g(\kappa) \to 0$, and the system is best described as an amorphous solid or glass.
\begin{figure}
\centerline{\includegraphics[height=2.5in]{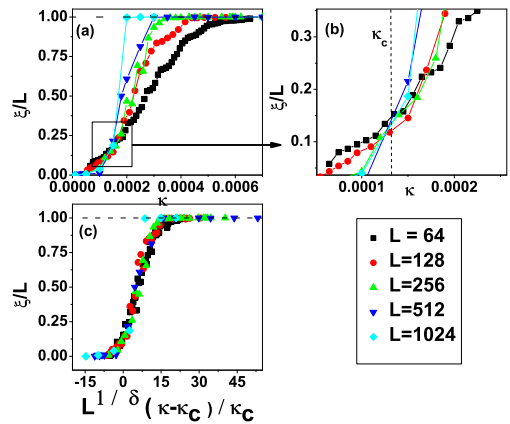}}
\caption{(color online) (a) The normalized most extended eigenwidth vs. interaction strength showing a crossing point at $\kappa \approx 0.00013$ in (b). (c) Finite size scaling analysis at data collapse for $1 / \delta \approx 0.42 \pm 0.05$.  }
\label{fig:1d}
\end{figure}

Typically in finite sized systems with quenched disorder, various quantities exhibit fluctuations that scale like $1/\sqrt L $. For smaller values of $L$, we can reduce the severity of these effects by performing averages over sufficiently large ensembles. However, at larger system sizes these quantities are self-averaging and one can reduce the number of required ensembles. Another source of numerical uncertainty is generated in the relaxation procedure outlined in  ref. \cite{1d_wc}, where residual forces can move the charges slightly out of equilibrium. Therefore, we have relaxed the charges to $\approx 10^{-12}$ in relative force magnitude.

The essence of the Anderson transition is that all of the eigenfunctions of some Hamiltonian, or $\mathbf{D(R - R')}$, containing an explicit form of randomness are localized if the normalized interaction strength, $\kappa$, is below some definite value $\kappa_c$. However, for $\kappa > \kappa_c$, some eigenmodes are localized in a transition that also depends on the value of the spectral variable $\omega^2$. That is, the eigenfunction spectrum is divided by a mobility edge, $\omega^{2}_c$ that separates the  frequency range in which all the states are localized from  the set of frequencies for which they are extended \cite{zimandisorder,anderson}. In this particular 1D WC system, the plasmon mobility edge has been confirmed \cite{mobility}. However, a transition involving $\kappa_c$ was not observed. We now report on this crucial aspect of the transition, as shown in Fig. \ref{fig:1d}. Our general strategy for determining the existence of such a transition is to diagonalize $\mathbf{D(R - R')}$, isolate the localization length $\xi$ of the most extended of all eigenmodes and observe how it behaves as a function of $\kappa$. We have computed $\xi$, normalized by the system size, $\xi(\kappa)/ L $. We require $\xi / L = 1$ for a truly extended eigenmode. As shown in Fig. \ref{fig:1d}(a), as $\kappa$ is decreased more states are localized until $\omega^{2}_c$ coalesces with the lower band edge, signifying complete localization. A bulk transition in the thermodynamic limit is confirmed by a common crossing point in the un-collapsed data shown in Fig. \ref{fig:1d}(b).
	
\begin{figure}
\centerline{\includegraphics[height=2.5in]{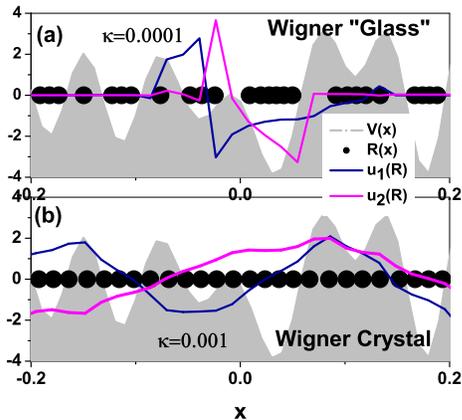}}
\caption{(color online) The distinct structural phases: The random potential $V(x)$, plotted together with relaxed equilibrium particle coordinates $R(x)$ and the two most extended eigenfunctions, $u_1 (R)$, $u_2 (R)$, for L=64. (a)$\kappa < \kappa_c$, in the Wigner "glass" phase, where the particle array has fractured into small domains that vibrate with localized plasmon eigenmodes. (b)$\kappa > \kappa_c$, Quasi-long range order with extended plasmon eigenmodes.}
\label{fig:glass}	
\end{figure}

Tests for a transition in the limit of an infinite system have been performed with the use of a finite-size scaling analysis, where we assumed the dependence of the principal quantities, $\xi /L = F\left( {L^{1/\delta } (\kappa  - \kappa _c )/\kappa _c } \right)$, in terms of a universal function $F$. We have verified that the data collapse is controlled by the single exponent $1 / \delta \approx 0.42 \pm 0.05$, of which has been optimized by applying a $\chi^2$ analysis in the overlapping regime yielding a $p$ value $= 0.40$, for which the data collapse can be interpreted as being in statistical agreement. For the specific parameters of our model, we have determined $\kappa_c \approx 0.00013 \pm 0.00002$.
\begin{figure}
\centerline{\includegraphics[height=2.5in]{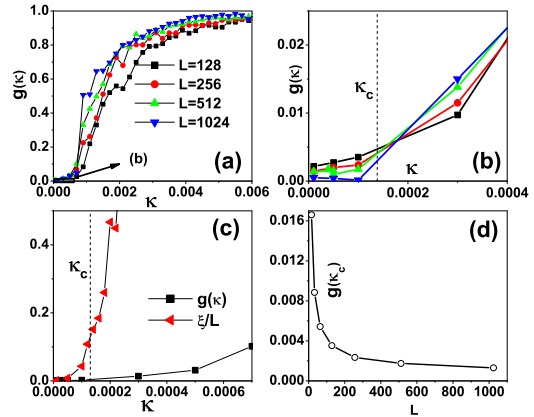}}
\caption{(color online) (a) $g(\kappa)$ at different system sizes. (b) A close up of $g(\kappa)$ in the amorphous regime where $g(\kappa = \kappa_c) \to 0$. (c) L=256, 1st Bragg peak height $g(\kappa)$ and most extended eigenwidth $\xi/L$, showing a coincidence of the structural and delocalization transitions at $\kappa = \kappa_c$.(d) $g(\kappa_c)$ as a function of system size $L$.}
\label{fig:peak_h}
\end{figure}
We have also expanded earlier investigations \cite{1d_wc} into the strong disorder regime near $\kappa_c$ and have observed $g(\kappa_c) \approx 0$ as shown in Fig.\ref{fig:peak_h}(a). Additionally, a focused plot of the critical regime is shown in Fig.\ref{fig:peak_h}(b), where $g(\kappa)$ reaches a lower saturation point precisely at $g(\kappa_c)$. We have verified that this is a bulk transition in the limit of a large system size as shown in Fig.\ref{fig:peak_h}(d). The disappearance of the first Bragg peak is a generic feature of structural phase transitions into amorphous, ``glasslike'' structures. A useful visual description is given by Fig.\ref{fig:glass}, where the relaxed particle configurations are plotted along with the random potential $V(x)$, in both the crystal($\kappa = 0.001$) and amorphous($\kappa = 0.0001$) regimes. Clearly, in the amorphous regime, where $\kappa < \kappa_c$, the charges tend to cluster into small crystallite domains inside of the potential wells. A characteristic length scale of these domains, $R_c$ tends like the effective capacity or width of the individual well. It is important that we note that $R_c$ is a disorder dependent quantity
and regarding previous work on similar systems, $R_c$ can be naively associated with the Larkin length \cite{larkin} for pinned elastic systems. However, it should be noted that we are studying a discrete model and by contrast the Larkin length arises as the equilibrium length in a force balancing scheme for a continuum model. Therefore, the association of $R_c$ with the Larkin length should be taken very loosely, if at all. The relationship between the plasmon delocalization transition and physics of pinning in elastic systems is not known.

We have confirmed a direct association of the quasi-long range crystalline order regime with the existence of a plasmon mobility edge by determining a coincidence of those respective transitions close to $\kappa = \kappa_c$. Upon inspection it is also evident from the two most extended eigenfunctions of the glassy regime, plotted as $u_1 (R)$ and $u_2 (R)$ in Fig. \ref{fig:glass}(a), that the localized plasmon eigenmodes tend to be confined near double well potentials. Consequently, the total data we have presented develops the basis for the phase diagram shown in Fig.\ref{fig:phase_d}.
\begin{figure}
\centerline{\includegraphics[height=2.5in]{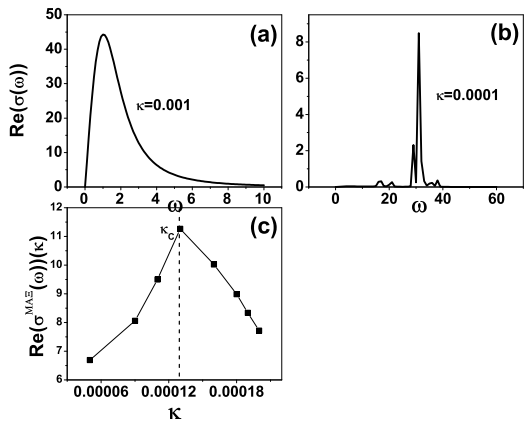}}
\caption{$Re[\sigma(\omega)]$ (L=128) (a) $\kappa > \kappa_c$ (b) $\kappa < \kappa_c$, (c) Predicted experimental signature of the transition. Largest spectral amplitude $Re[\sigma_{MAX}(\omega)]$ forms a local maximum precisely at $\kappa_c$. }
\label{fig:ac_cond}
\end{figure}

Let us now focus on the behavior of the $\sigma(\omega)$ in the regimes separated by $\kappa_c$. From the definition of $\sigma(\omega)$ given earlier, $Re[\sigma(\omega)]\propto Im[i \omega G(\omega)] \propto \rho(\omega)$, where $\rho(\omega)$ is the plasmon spectral density. It is well known from previous studies of Anderson localization that $\rho(\omega)$ in the strictly localized regime exhibits, discrete and well separated peaks that correspond to the bound state spectrum\cite{zimandisorder}. We computed $Re[\sigma(\omega)]$ for both $\kappa < \kappa_c$ and $\kappa > \kappa_c$ as shown in Fig. \ref{fig:ac_cond}. Clearly in the localized regime, Fig. \ref{fig:ac_cond} (b) we recover the expected discreteness, essential to Anderson localization. It follows that in the thermodynamic limit, the smaller individual peaks would smear out into a continuum. However, this differs from the $\kappa > \kappa_c$ response of Fig. \ref{fig:ac_cond} (a), corresponding to a large, single and intact domain. Therefore we emphasize that an experimental signature of the plasmon Anderson transition would be present in the frequency dependent AC conductivity as a shift between these two general forms. We have examined the maximum amplitude which is defined as the maximum value of $Re[\sigma(\omega)]$ for a full spectra, as a function of $\kappa$. Evidently, a local maximum forms precisely at $\kappa_c$ as shown in Fig. \ref{fig:ac_cond} (c).

Lastly, we mention that $\sigma(\omega)$ for a 1D elastic system with short ranged interactions, pinned by disorder was studied in \cite{fogler}, where an elastic string was pinned into small domains with effective lengths that define the Larkin scale $R_c$. If we consider the AC response of the individual domain it has been predicted that the peak frequency $\omega_p$ should scale with the domain size $\omega_p \sim  1/R_c$\cite{fogler}. This dependence appears consistent with the shift into a spread of larger $\omega_p$ values as shown in Fig. \ref{fig:ac_cond} (b). Intuitively, one would expect a smaller domain to vibrate with a higher $\omega_p$ values.

We conclude by noting that the results of this paper have direct relevance to not only 1DWC systems but the general properties of bosonic excitations in random media and amorphous solids\cite{bragg,gurarie}. The phase diagram shown in Fig. \ref{fig:phase_d} is the primary means for linking a system's  structural properties to the localization of its collective modes. A more general topic that should be pursued further is the study of delocalization transitions in harmonic oscillator systems in any dimensionality that have a non-crystalline configuration at equilibrium. Is there a critical power of the interactions between the oscillators that determines whether the model is critical or not? For example we have considered a $1/R$ Coulomb potential in 1D. Thus, for a general power law interaction $1/R^{\alpha}$ in any dimension $d$, what is the critical value of $\alpha$ or $d$ for which a $\kappa_c$ should exist? This suggests the possibility of multiple parameter scaling features (including $N$) with a more intricate phase diagram. A broader motivation can be extended recent developments in biological and optical lattice systems. Another important avenue to explore further is the role of the double well potential and the various tunneling and energy splitting processes. Perhaps the low temperature, universal properties associated with the double well potential as generalized by Anderson-Halperin-Varma theory of amorphous solids plays a role in describing the plasmon conductance of the glassy regime\cite{halperin}. Many issues regarding quantum mechanical effects such as dissipation and the role of plasmon exchange and tunneling through finite barriers will also be addressed in future investigations.

We thank Professors G. Gruner, S.E. Brown, and S. Chakravarty for useful discussions.
J.R. Acknowledges support of the NSF through grant no. DMR 04-04507.


\begin{thebibliography}{17}
\expandafter\ifx\csname natexlab\endcsname\relax\def\natexlab#1{#1}\fi
\expandafter\ifx\csname bibnamefont\endcsname\relax
  \def\bibnamefont#1{#1}\fi
\expandafter\ifx\csname bibfnamefont\endcsname\relax
  \def\bibfnamefont#1{#1}\fi
\expandafter\ifx\csname citenamefont\endcsname\relax
  \def\citenamefont#1{#1}\fi
\expandafter\ifx\csname url\endcsname\relax
  \def\url#1{\texttt{#1}}\fi
\expandafter\ifx\csname urlprefix\endcsname\relax\def\urlprefix{URL }\fi
\providecommand{\bibinfo}[2]{#2}
\providecommand{\eprint}[2][]{\url{#2}}

\bibitem[{\citenamefont{Matveev}(2004)}]{matveev}
\bibinfo{author}{\bibfnamefont{K.~A.} \bibnamefont{Matveev}},
  \bibinfo{journal}{Phys. Rev. Lett.} \textbf{\bibinfo{volume}{92}},
  \bibinfo{eid}{106801} (\bibinfo{year}{2004}).

\bibitem[{\citenamefont{Schulz}(1993)}]{schulzwigner}
\bibinfo{author}{\bibfnamefont{H.~J.} \bibnamefont{Schulz}},
  \bibinfo{journal}{Phys. Rev. Lett.} \textbf{\bibinfo{volume}{71}},
  \bibinfo{pages}{1864} (\bibinfo{year}{1993}).

\bibitem[{\citenamefont{Piacente et~al.}(2004)\citenamefont{Piacente,
  Schweigert, Betouras, and Peeters}}]{peeters}
\bibinfo{author}{\bibfnamefont{G.}~\bibnamefont{Piacente}},
  \bibinfo{author}{\bibfnamefont{I.~V.} \bibnamefont{Schweigert}},
  \bibinfo{author}{\bibfnamefont{J.~J.} \bibnamefont{Betouras}},
  \bibnamefont{and} \bibinfo{author}{\bibfnamefont{F.~M.}
  \bibnamefont{Peeters}}, \bibinfo{journal}{Phys. Rev. B}
  \textbf{\bibinfo{volume}{69}}, \bibinfo{pages}{045324}
  (\bibinfo{year}{2004}).

\bibitem[{\citenamefont{Klironomos et~al.}(2005)\citenamefont{Klironomos,
  Ramazashvili, and Matveev}}]{qwire}
\bibinfo{author}{\bibfnamefont{A.~D.} \bibnamefont{Klironomos}},
  \bibinfo{author}{\bibfnamefont{R.~R.} \bibnamefont{Ramazashvili}},
  \bibnamefont{and} \bibinfo{author}{\bibfnamefont{K.~A.}
  \bibnamefont{Matveev}}, \bibinfo{journal}{Phys. Rev. B}
  \textbf{\bibinfo{volume}{72}}, \bibinfo{eid}{195343}
  (pages~\bibinfo{numpages}{5}) (\bibinfo{year}{2005}).

\bibitem[{\citenamefont{Hiraki and Kanoda}(1998)}]{jap_exp}
\bibinfo{author}{\bibfnamefont{K.}~\bibnamefont{Hiraki}} \bibnamefont{and}
  \bibinfo{author}{\bibfnamefont{K.}~\bibnamefont{Kanoda}},
  \bibinfo{journal}{Phys. Rev. Lett.} \textbf{\bibinfo{volume}{80}},
  \bibinfo{pages}{4737} (\bibinfo{year}{1998}).

\bibitem[{\citenamefont{Ziman}(1979)}]{zimandisorder}
\bibinfo{author}{\bibfnamefont{J.~M.} \bibnamefont{Ziman}},
  \emph{\bibinfo{title}{Models of Disorder}} (\bibinfo{publisher}{Cambridge
  University Press}, \bibinfo{address}{Cambridge}, \bibinfo{year}{1979}),
  chap.~\bibinfo{chapter}{9}.

\bibitem[{\citenamefont{Anderson}(1958)}]{anderson}
\bibinfo{author}{\bibfnamefont{P.~W.} \bibnamefont{Anderson}},
  \bibinfo{journal}{Phys. Rev.} \textbf{\bibinfo{volume}{109}},
  \bibinfo{pages}{1492} (\bibinfo{year}{1958}).

\bibitem[{\citenamefont{Akhanjee and Rudnick}(2007{\natexlab{a}})}]{mobility}
\bibinfo{author}{\bibfnamefont{S.}~\bibnamefont{Akhanjee}} \bibnamefont{and}
  \bibinfo{author}{\bibfnamefont{J.}~\bibnamefont{Rudnick}},
  \bibinfo{journal}{Phys. Rev. B} \textbf{\bibinfo{volume}{75}},
  \bibinfo{eid}{012302} (\bibinfo{year}{2007}{\natexlab{a}}).

\bibitem[{\citenamefont{Akhanjee and Rudnick}(2007{\natexlab{b}})}]{1d_wc}
\bibinfo{author}{\bibfnamefont{S.}~\bibnamefont{Akhanjee}} \bibnamefont{and}
  \bibinfo{author}{\bibfnamefont{J.}~\bibnamefont{Rudnick}},
  \bibinfo{journal}{cond-mat/0612131, To Appear in Phys. Rev. B}
  \textbf{\bibinfo{volume}{76}} (\bibinfo{year}{2007}{\natexlab{b}}).

\bibitem[{\citenamefont{Chitra et~al.}(2002)\citenamefont{Chitra, Giamarchi,
  and Doussal}}]{giampwc}
\bibinfo{author}{\bibfnamefont{R.}~\bibnamefont{Chitra}},
  \bibinfo{author}{\bibfnamefont{T.}~\bibnamefont{Giamarchi}},
  \bibnamefont{and} \bibinfo{author}{\bibfnamefont{P.~L.}
  \bibnamefont{Doussal}}, \bibinfo{journal}{Phys. Rev. B}
  \textbf{\bibinfo{volume}{65}} (\bibinfo{year}{2002}).

\bibitem[{\citenamefont{Dobrosavljevi\ifmmode~\acute{c}\else \'{c}\fi{}
  et~al.}(2003)\citenamefont{Dobrosavljevi\ifmmode~\acute{c}\else \'{c}\fi{},
  Tanaskovi\ifmmode~\acute{c}\else \'{c}\fi{}, and Pastor}}]{pastor}
\bibinfo{author}{\bibfnamefont{V.}~\bibnamefont{Dobrosavljevi\ifmmode~\acute{c%
}\else \'{c}\fi{}}},
  \bibinfo{author}{\bibfnamefont{D.}~\bibnamefont{Tanaskovi\ifmmode~\acute{c}\%
else \'{c}\fi{}}}, \bibnamefont{and} \bibinfo{author}{\bibfnamefont{A.~A.}
  \bibnamefont{Pastor}}, \bibinfo{journal}{Phys. Rev. Lett.}
  \textbf{\bibinfo{volume}{90}}, \bibinfo{pages}{016402}
  (\bibinfo{year}{2003}).

\bibitem[{\citenamefont{Ashcroft and Mermin}(1976)}]{AM}
\bibinfo{author}{\bibfnamefont{N.~W.} \bibnamefont{Ashcroft}} \bibnamefont{and}
  \bibinfo{author}{\bibfnamefont{N.~D.} \bibnamefont{Mermin}},
  \emph{\bibinfo{title}{Solid state physics}} (\bibinfo{publisher}{Holt
  Rinehart and Winston}, \bibinfo{address}{New York}, \bibinfo{year}{1976}).

\bibitem[{\citenamefont{Larkin and Ovchinnikov}(1979)}]{larkin}
\bibinfo{author}{\bibfnamefont{A.}~\bibnamefont{Larkin}} \bibnamefont{and}
  \bibinfo{author}{\bibfnamefont{Y.}~\bibnamefont{Ovchinnikov}},
  \bibinfo{journal}{J. Low Temp. Phys.} \textbf{\bibinfo{volume}{34}},
  \bibinfo{pages}{409} (\bibinfo{year}{1979}).

\bibitem[{\citenamefont{Fogler}(2002)}]{fogler}
\bibinfo{author}{\bibfnamefont{M.~M.} \bibnamefont{Fogler}},
  \bibinfo{journal}{Phys. Rev. Lett.} \textbf{\bibinfo{volume}{88}},
  \bibinfo{pages}{186402} (\bibinfo{year}{2002}).

\bibitem[{\citenamefont{Giamarchi and Le~Doussal}(1994)}]{bragg}
\bibinfo{author}{\bibfnamefont{T.}~\bibnamefont{Giamarchi}} \bibnamefont{and}
  \bibinfo{author}{\bibfnamefont{P.}~\bibnamefont{Le~Doussal}},
  \bibinfo{journal}{Phys. Rev. Lett.} \textbf{\bibinfo{volume}{72}},
  \bibinfo{pages}{1530} (\bibinfo{year}{1994}).

\bibitem[{\citenamefont{Gurarie and Altland}(2005)}]{gurarie}
\bibinfo{author}{\bibfnamefont{V.}~\bibnamefont{Gurarie}} \bibnamefont{and}
  \bibinfo{author}{\bibfnamefont{A.}~\bibnamefont{Altland}},
  \bibinfo{journal}{Phys. Rev. Lett.} \textbf{\bibinfo{volume}{94}},
  \bibinfo{eid}{245502} (\bibinfo{year}{2005}).

\bibitem[{\citenamefont{Anderson et~al.}(1972)\citenamefont{Anderson, Halperin,
  and Varma}}]{halperin}
\bibinfo{author}{\bibfnamefont{P.}~\bibnamefont{Anderson}},
  \bibinfo{author}{\bibfnamefont{B.~I.} \bibnamefont{Halperin}},
  \bibnamefont{and} \bibinfo{author}{\bibfnamefont{C.~M.} \bibnamefont{Varma}},
  \bibinfo{journal}{Phil. Mag.} \textbf{\bibinfo{volume}{25}},
  \bibinfo{pages}{1} (\bibinfo{year}{1972}).

\end{thebibliography}
\end{document}